\input amssym.def
\input amssym
\newcommand{\Beq}{\begin{equation}}
\newcommand{\Eeq}{\end{equation}}
\newcommand{\Beqa}{\begin{eqnarray}}
\newcommand{\Eeqa}{\end{eqnarray}}
\newcommand{\End}{\nonumber\\}
\newcommand{\Lba}{\bigl(}
\newcommand{\Lbb}{\Bigl(}
\newcommand{\Lbc}{\biggl(}

\newcommand{\Rba}{\bigr)}
\newcommand{\Rbb}{\Bigr)}
\newcommand{\Rbc}{\biggr)}

\newcommand{\Lsa}{\bigl[}
\newcommand{\Rsa}{\bigr]}
\newcommand{\Lsb}{\Bigl[}
\newcommand{\Rsb}{\Bigr]}
\newcommand{\Lsc}{\biggl[}
\newcommand{\Rsc}{\biggr]}
\newcommand{\Real}{\Bbc{R}}
\newcommand{\Bbc}[1]{{\Bbb{#1}}}
\newcommand{\Bmn}[2]{{\Bbc{B}}\sp {#1,#2}}
\newcommand{\Rowrow}[4]{#1\sp 1,\dots,#1\sp {#2},#3\sp 1,\dots,#3\sp {#4}}
\newcommand{\Row}[2]{#1\sp 1,\dots,#1\sp {#2}}
\newcommand{\Summun}{\sum_{\mu\in M_n}}
\newcommand{\Ginf}{G\sp {\infty}(n)}
\newcommand{\Bare}{{\Bbc{B}}}
\newcommand{\Gr}{${\cal G}$-}
\newcommand{\Cinf}[1]{#1\sp {\infty}}
\newcommand{\Gws}[2]{\Lba \Bare\sp {#1,#2}\Rba\sp {I}}
\newcommand{\Spath}[1]{\underline{\theta}(#1),\underline{\rho}(#1)}
\newcommand{\Tord}{0\leq t_1 < \dots < t_N \leq t}
\newcommand{\Rowf}[1]{\underline{\theta}\sp 1,\underline{\rho}\sp 1, \dots
,\underline{\theta}\sp N\underline{\rho}\sp N}
\newcommand{\Measf}{\exp\Lsb -
\Lba\underline{\rho}\sp 1.\underline{\theta}\sp 1 + \underline{\rho}\sp 2.
(\underline{\theta}\sp 2-\underline{\theta}\sp 1) + \dots +
\underline{\rho}\sp N(\underline{\theta}\sp N -
\underline{\theta}\sp {N-1})\Rba\Rsb}
\newcommand{\Half}{{\textstyle\frac{1}{2}}}
\newcommand{\Ito}{It\\sp o}
\newcommand{\Ebold}{{\Bbc E}}
\newcommand{\Edef}{=_{{\rm def}}}
\newcommand{\Stochint}[3]{#1\sp 1_{#3},\dots, #1\sp {#2}_{#3}}
\newcommand{\Comp}{{\Bbc C}}
\newcommand{\Eqe}{=_{\Ebold}}
\newcommand{\Con}[3]{\Gamma\sp {#3}_{#1#2}}
\newcommand{\Cog}[3]{A_{#1#2}\sp #3}
\newcommand{\Lbe}{\Half(d + \delta)\sp 2}
\newcommand{\Wbk}{Weitzenbock}
\newcommand{\Del}[1]{\delta_{\theta\sp {#1}}}
\newcommand{\Det}[1]{\delta_{\eta\sp {#1}}}
\newcommand{\Cur}[4]{R_{#1#2#3}{}\sp {#4}}
\newcommand{\Cuf}[4]{F_{#1#2#3}{}\sp {#4}}
\newcommand{\Frac}[2]{{\textstyle\frac{#1}{#2}}}
\newcommand{\trace}{{\rm tr}}
\newcommand{\Trace}{{\rm Tr}}

\newcommand{\Indint}{\trace \exp\Lba \frac{- F}{2\pi} \Rba \det \Lbb
\frac{i\Omega/2\pi}{\tanh i\Omega/2\pi} \Rbb\sp {\Half}}
\newcommand{\Sumsum}{\sum_{\mu \in M_n} \sum_{p=1}\sp n}
\newcommand{\Hot}{{\rm h.o.t.}}
\newcommand{\Factor}[1]{(2\pi#1)\sp {-\Frac m2}}
\newcommand{\One}[1]{#1\sp 1{}}
\newcommand{\Two}[1]{#1\sp 2{}}
\newcommand{\Som}{S(O(M),E)}
\newcommand{\Etah}[1]{\hat{\eta}\sp {#1}}

\newcommand{\Curv}[2]{\Omega_{#1}{}\sp {#2}}
\newcommand{\Phitt}[1]{{\phi\sp {#1}\over\sqrt {2\pi t}}}
\newcommand{\Xhat}[1]{\hat{x}\sp {#1}}
\title{Anticommuting Variables, Fermionic Path Integrals and
Supersymmetry\\
Lectures given at the XXVIII Karpacz Winter School of Theoretical Physics,
Poland 1992\\
KCL-TH-92-5}
\author{Alice Rogers\thanks{Supported by the Royal Society}\\
Department of Mathematics\\
King's College\\
Strand, London WC2R 2LS}
\date{June 1992}
\documentstyle[12pt]{article}
\begin{document}
\bibliographystyle{plain}
\maketitle
\begin{abstract}
Fermionic Brownian paths are defined as paths in a space
para\-metr\-ised by anticommuting variables. Stochastic calculus for these
paths, in conjunction with classical Brownian paths, is described;
Brownian paths on supermanifolds are developed and applied to establish a
Feynman-Kac formula for the twisted Laplace-Beltrami operator on
differential forms taking values in a vector bundle. This formula is used
to give a proof of the Atiyah-Singer index theorem which is rigorous while
being closely modelled on the supersymmetric proofs in the physics
literature. \end{abstract}
\section{Introduction}
These lectures concern a generalisation of Brownian paths to
include
ferm\-ion\-ic paths, which are paths in spaces parametrised by
anti\-commut\-ing variables. The aim of this work is to
provide a
rigorous
version of some heuristic constructions used with great effect
in
quantum physics, and in application to geometry.

While fermionic paths do not directly model any physical
quantity, they
provide a technique for investigating differential operators
on spaces
of functions of anticommuting variables; since one can obtain
an
analogue of the Schr\"odinger representation for fermion
operators
in quantum physics using just this kind of operator, such
paths are
useful in path integral quantisation of theories with
fermions, as was
first observed in a highly original paper of Martin
\cite{Martin}.
Additionally, functions on carefully constructed
supermanifolds are
equivalent to differential forms or spinor fields, which can
then be
analysed by the geometric fermionic path integration
techniques
described in these lectures. It is of course possible to
handle
fermionic quantization without using
anticommuting variables; however, in the author's opinion,
they are a
valuable aid to intuition, particularly in supersymmetric
models, and
these lectures are presented in the hope of showing that such
variables
also have analytic power.

Section 2 of these lectures introduces fermionic Brownian
paths, and the
corresponding Wiener measure; a brief review of conventional
stochastic
calculus and its use in deriving Feynman-Kac (or path
integral) formulae
for diffusion operators is then given. In section 4 it is
shown how
these methods may be extended to include fermionic paths.
Section 5
reviews the standard construction of Brownian paths on
manifolds, and
extends the construction to paths on carefully chosen
supermanifolds,
leading to a Feynman-Kac formula for the Laplace-Beltrami
operator on
twisted differential forms. In the final section these
techniques are
applied to give a rigorous version of the supersymmetric
proofs of the
Atiyah-Singer index theorem. A more formal account of this
work, with
analytic details, may be found in \cite{SCSTWO}.

The following conventions will be used: even variables, which
anticommute with all variables, will be denoted by lower case
latin
letters, while odd variables, which anticommute with one
another but
commute with even variables, will be denoted by lower case
greek
letters. $\Bmn mn$ will denote the space whose elements are
$m+n$-tuples
$(\Rowrow xm{\theta}n)$ with $\Row xm$ even and
$\Row{\theta}n$ odd. The
space  of functions of $n$ odd variables $\Row{\theta}n$ of
the form
\Beq
f(\Row{\theta}n) = \Summun
f_{\mu}\theta\sp {\mu_1}\dots\theta\sp {\mu_k}
\label{THETAEXP}\Eeq
where each $\mu=\mu_1\dots\mu_k$ is a multi-index, with
$1<\mu_1\leq\dots\leq\mu_k<n$, and $M_n$ denotes the set of
all such
multi-indices (including the empty one), will be denoted
$\Ginf$. The
coefficients $f_{\mu}$ will take values in some specified
space.
Integration of functions of anticommuting variables will
follow the
Berezin prescription \cite{Berezin1}
\Beq
\int d\sp n\theta \, f(\theta) = f_{1\dots n}
\Eeq
where $f_{1\dots n}$ is the coefficient of
$\theta\sp 1\dots\theta\sp n$ in
the expansion (\ref{THETAEXP}) of the function $f$. The
integral kernel
of an operator $H$ on the function space $\Ginf$ is a function
$H(\Rowrow{\theta}n{\phi}n)$ of $2n$ anticommuting variables
such that
\Beq
Hf(\theta) = \int d\sp n\phi H(\theta,\phi) f(\phi).
\Eeq
A useful feature of the Berezin integral is that it allows the
trace of
an operator to be calculated from its kernel; it can be shown
by
explicit calculation that
\Beq
\trace H = \int d\sp n \theta H(\theta,-\theta).
\label{TRACE}
\Eeq
Through out these lectures the language of probability theory
will be
used for fermionic analogues; however such analogues do not
have all the
properties of their classical counterparts. For instance, the
Grassmann
Wiener measure defined in the next section is not a true
measure, but it
seems useful to use the same terminology, indicating the
departure from
convention by the prefix ${\cal G}$, so that, for example,
Grassmann
Wiener measure is said to be a \Gr measure.
%
%
\section{Fermionic Brownian paths}
The fermionic analogue of Brownian paths and Wiener measure
will now be
defined. Letting $I$ denote the closed interval $[0,t]$ of the
real
line, Grassmann Wiener measure is a \Gr measure on
$\Gws0{2n}$, the
space
of paths in $2n$-\break
dimensional anticommuting space. A typical element of
this space is denoted\break
$(\theta\sp 1(t), \dots, \theta\sp n(t), \rho\sp 1(t),\dots,\rho\sp n(t)|t
\in I)$ or
$(\Spath t|t \in I)$. The \Gr measure is then defined by
specifying its
finite distributions. That is, suppose that $G$ is a function
on
$\Gws0{2n}$ which actually only depends on $\Spath{t_i}$ at a
finite set
of times $t_i,i=1,\dots,N$ with $\Tord$, so that
\Beq
G(\Spath t) = G(\Spath{t_1},\dots ,\Spath{t_N}).
\Eeq
Then
\Beqa
\Ebold [G] &\Edef& \int d\mu_F \, G \End
&\Edef& \int d\sp {nN}\theta\, d\sp {nN}\rho\, F_N(\Rowf N) G(\Rowf N)
\Eeqa
where
\Beq
F_N(\Rowf N) = \Measf
\label{FWM}\Eeq
with
\Beq
\underline{\rho}\sp r.\underline{\theta}\sp r = \sum_{i=1}\sp n
\rho\sp {ir}\theta\sp {ir}
\Eeq
for $r = 1, \dots, N$. The distribution $F_N$ corresponds to
the
heuristic fermionic path integral measure $\exp \int_0\sp t
\overline{\psi}(s)\dot{\psi}(s) \, ds$. The distributions
$F_N$ all have
weight one, so that they are probability distributions. They
also obey
the consistency condition
\Beq
\int d\sp n\theta\sp r\, d\sp n\rho\sp r \, F_N(\Rowf N) = F_{N-1}
(\underline{\theta}\sp 1,\underline{\rho}\sp 1 , \dots
,\hat{\underline{\theta}\sp r},\hat{\underline{\rho}\sp r}, \dots,
\underline{\theta}\sp N\underline{\rho}\sp N),
\Eeq
where the caret indicates omission of an argument.
Functions such as $G$ above are called \Gr random variables.
Particular
examples are $\theta_s=\theta\sp i(s)$ and $\rho_s=\rho\sp i(s)$ for
some $i,
0 \leq i \leq n$
and some $s, 0 < s < t$. The collection
$\{\theta\sp i_s,\rho\sp i_s| i= 1,
\dots, n, s \in I\}$ is called fermionic Brownian motion. More
complicated
\Gr random variables, such as $\int_0\sp t V(\theta_s,\rho_s)
ds$, can be defined by
a limiting process. One has the Feynman-Kac formula
\Beq
(\exp -Ht) f(\theta) = \Ebold \Lsa \exp \Lba-\int_0\sp t
V_{\mu\nu}\theta_s\sp {\mu} (i\rho_s)\sp {\nu} ds\Rba f(\theta +
\theta_t)
\Rsa
\label{FFK}\Eeq
where $H = \sum_{\mu,\nu \in M_n}V_{\mu\nu}
\theta\sp {\mu}(\frac{\partial}{\partial\theta})\sp {\nu}$
\cite{GBM}. (Note that the
free fermionic Hamiltonian is zero.)

This measure can be combined as a direct product with
conventional
Wiener measure on $\Real\sp m$ to give a measure on the space
$\Gws m{2n}$
of paths in the superspace $\Bmn m{2n}$.
\section{Stochastic calculus}
Heuristic derivations of path integral formulae meet greater
difficulties when considering (even in a purely bosonic
setting)
Hamiltonians of the form
\Beq
H = -\Half g\sp {ij}(x)\partial_1\partial_j + h\sp i(x)\partial_i +
V(x).
\label{LAP}\Eeq
It is possible to handle the direct, time-slicing approach to
such
Hamiltonians in an analytically rigorous way, but this
approach is not
always
useful, because it requires a knowledge of the very operator
one is
hoping to study, or of a closely related operator. Much more
effective
are the techniques of stochastic calculus. These are
unfamiliar to many
physicists, and so a summary of those aspects of the standard
theory
which are important in  applications to diffusions (or
imaginary-time
Schr\"odinger equations) will now be given, before showing how
these
methods may be extended to fermionic Brownian motion.

Brownian paths $(b_s|s\in I)$ are almost nowhere smooth, but
nevertheless sufficiently regular for $\int_0\sp t f_s db_s$ to
be defined
in the following manner:
let $f_s = f(\{b_u|u \leq s\})$ with
\Beq
\Ebold\Lsb \int_0\sp t |f_s|\sp 2\, ds \Rsb< \infty.
\Eeq
(This is a rather loose
definition of an
adapted stochastic process on Wiener space.)
Then
\Beq
\int_0\sp t f_s db_s = \lim_{N\to\infty} \sum_{r=0}\sp {2\sp N-1}
f_{t_r}
(b_{t_{r+1}} - b_{t_r}),
\Eeq
where $t_r = rt/2\sp N$. \Ito\ integrals with respect to multi-
dimensional
Brownian motion may be defined in a similar manner. (A fuller
account of this may be found in a number of places; for
physicists the work of Simon is an accessible account.)

A crucial formula is the \Ito\ formula for the change of
variable.
Suppose that $\Stochint apt$ are stochastic integrals on the
Wiener
space of Brownian
paths in $\Real\sp m$ over the time interval $I$; that is, there
exist
adapted stochastic processes $f\sp i_{a,s},g\sp i_{s}, i=1, \dots p,
a=1,
\dots m, s \in I$ and random variables $a\sp i_0, i = 1, \dots
,p$ such
that
\Beq
a\sp i_s = a\sp i_0 + \int_0\sp s \Lsa\sum_{a=1}\sp mf\sp i_{a,u} db\sp a_u
+g\sp i_s ds
\Rsa.
\Eeq
Then, if $F$ is a sufficiently regular function of $p$
variables,
$F(\Stochint apt)$ is also a stochastic integral and
\Beqa
F(\Stochint aps) - &F&(\Stochint ap0) \End
&=& \int_0\sp s \sum_{i=1}\sp p\sum_{a=1}\sp m \partial_iF(\Stochint
apu)(f\sp i_{a,u} + g\sp i_u) \, du \End
& & + \Half \sum_{i,j=1}\sp p \sum_{a=1}\sp m \partial_i\partial_jF
(\Stochint apu)f\sp i_{a,u}f\sp j_{a,u} ds.
\label{ITO}\Eeqa
This formula is proved much as the corresponding formula in
conventional
calculus is proved; the key estimate is that
\Beq
\Ebold\Lsb(b\sp a_{s+\delta s}-b\sp a_s)(b\sp c_{s+\delta s}-
b_s\sp c)\Rsb = \Half
\delta\sp {ac} \delta s,
\Eeq
which accounts for the presence of the second order term in
(\ref{ITO}).

This formula will now be applied to obtain a path-integral
expression
for $\exp(-Ht)$ when $H$ is a second order elliptic operator
on
$\Real\sp m$ of the form (\ref{LAP}). First suppose that
functions $e\sp i_a(x), i=1,\dots,m, a= 1, \dots ,p$ satisfy
\Beq
 e\sp i_a(x)e\sp j_a(x) = g\sp {ij}(x).
\Eeq
(Here and in the remainder of the paper the summation
convention that
repeated indices are to be summed over their range will be
used.) Then
consider the stochastic differential equation
\Beq
x\sp i_s = x\sp i + \int_0\sp s \Lba e\sp i_a(x_u) db\sp a_u - h\sp i(x_u)
du
\Rba
\Eeq
where $x\sp i \in \Real\sp m$. (Such equations are known to have
unique
solutions $x_s$, provided that the functions $g\sp {ij}$ and
$h\sp i$ are
sufficiently
regular.) Given $f\in\Cinf C$, set
\Beq
F_s(f,x) = \exp\Lsb-\int_0\sp s V(x_u)du\Rsb f(x_s).
\Eeq
Then, applying the \Ito\ formula (\ref{ITO}), one obtains
\Beqa
F_s(f,x)-F_0(f,x) &=& \int_0\sp s \Lsc \exp\Lsb -\int_0\sp u V(x_v)
dv \Rsb
\End
\times \Lbb -V(x_u) f(x_u)du &+& \partial_if(x_u)\Lba
e\sp i_a(x_u)db\sp a_u -
h\sp i(x_u) du \End
 &+& \Half \partial_i\partial_j e\sp i_a(x_u) e\sp j_a(x_u)du \Rba
\Rbb\Rsc.
\Eeqa
If the operator $U_s$ on $\Cinf C(\Real\sp m)$ is now defined by
setting
\Beq
U_s f(x) = \Ebold(F_s(f,x),
\Eeq
then (using the fact that the expectation of the \Ito\
integral of an
adapted function is always zero), one finds that
\Beq
U_s f(x) - f(x) = \int_0\sp s U_u H f(x) du
\Eeq
where $H$ is the operator
\Beq
H = -\Half g\sp {ij}(x)\partial_1\partial_j + h\sp i(x)\partial_i +
V(x),
\Eeq
and hence
\Beq
U_s = \exp -Hs,
\Eeq
so that the Feynman-Kac formula
\Beq
(\exp-Hs)f(x) = \exp\Lsb-\int_0\sp s V(x_u)du\Rsb f(x_s)
\Eeq
has been established.
\section{Fermionic paths and stochastic calculus}
Fermionic paths can be incorporated into stochastic calculus,
but it is
neither necessary nor possible to define integrals along
fermionic
paths. In the case of standard, bosonic Brownian paths,
stochastic
integrals allow path integral quantization techniques to
handle
Hamiltonians which are arbitrary second-order elliptic
operators of the
form (\ref{LAP}); in the case of fermionic paths, because they
are
defined in phase space, all derivative operators can be
handled by the
simple Feynman-Kac formula (\ref{FFK}), without the necessity
of
introducing stochastic integrals. Moreover, it can be seen in
a number
of ways that fermion paths are too irregular to allow any
simple
analogue of the \Ito\ integral; inspection of the Fourier mode
analysis
of fermionic Brownian paths in \cite{GBM} indicates that their
derivatives would be divergent, while the lack of explicit
time-dependence in the fermionic Wiener distributions
(\ref{FWM}) means
that increments of all orders remain of order $1$, and thus no
analogue
of the \Ito\ formula (\ref{ITO}) exists for fermions unless
one includes
derivatives of all orders.

However fermionic paths can be included in the integrand of
bosonic
\Ito\ integrals, with the corresponding \Ito\ formula being
that, if
\Beq
Z\sp i_s = Z\sp i + k(\theta_s) + \int_0\sp s
h\sp i(Z_u,\theta_u,\rho_u)du +
\int_0\sp t e\sp i_a(Z_u,\theta_u,\rho_u)db\sp a_u
\Eeq
then, for sufficiently regular functions $F$,
\Beqa
F(Z_s)-F(Z_0) &\Eqe& \int_0\sp u \Lsb\Lba
h\sp i(Z_u,\theta_u,\rho_u) +
e\sp i_a(Z_u,\theta_u,\rho_u)db_u\sp a\Rba \partial_i F(Z_u) \End
& &+ \Half
e\sp i_j(Z_u,\theta_u,\rho_u)e\sp j_a(Z_u,\theta_u,\rho_u)
\partial_i\partial_j F(Z_u) du \Rsb,
\Eeqa
where the symbol $\Eqe$ indicates that two \Gr random
variables have
equal expectations. This leads to a Feynam-Kac formula for
Hamiltonians
which resemble
(\ref{LAP}) but include fermion operators. Full details may be
found in
\cite{SCSONE}.
\section{Brownian paths on supermanifolds}
In this section the approach to Brownian paths on manifolds
which may be
found in the work of Elworthy \cite{Elwort}, Malliavin
\cite{Mallia}
and Ikeda and Watanabe \cite{IkeWat} will be briefly reviewed
and then
extended to Brownian paths on
supermanifolds.

The theory of geometric Brownian paths on Riemannian manifolds
has two
components. First,
suppose that $V_a,a=1, \dots ,p$ are vector fields on an
$m$-dimensional manifold $M$. Then, in local coordinates $x\sp i,
i=1,
\dots ,m$ on some coordinate patch of $M$, $V_a = V_a\sp i(x)
(\partial/\partial x\sp i)$. If one considers the stochastic
differential
equations
\Beq
x_s\sp i = x\sp i + \int_0\sp s \Lsa V\sp i_a(x_u) db\sp a_u + \Half
V_a\sp j(x_u)
\partial_j V_a\sp i(x_u) du \Rsa,
\label{ELW}\Eeq
(with $b\sp a$ being $p$-dimensional Brwonian motion), one finds
that
under change of coordinate $x\sp i \to \tilde{x}\sp i(x)$ this
stochastic
differential equation transforms covariantly -- the non-
tensorial part
of the equation exactly compensates for the second order term
in the
\Ito\ formula for change of variable. The Stratonovich
integral allows
one to systemise this; given two stochastic integrals $X$ and
$Y$ with
\Beqa
X_s &=& \int_0\sp s \Lsa f_{a,u} db\sp a_u + f_{0,u} du \Rsa,\End
Y_s &=& \int_0\sp s \Lsa g_{a,u} db\sp a_u + g_{0,u} du \Rsa,
\Eeqa
the Stratonovich differential is defined by setting
\Beq
Y_s\circ dX_s =_{def} Y_s(f_{a,s} db\sp a_s + f_{0,s} ds)+ \Half
f_{a,s}
g_{a,s} ds.
\Eeq
The integrand in (\ref{ELW}) can then be expressed as
$V\sp i_a(x_u) \circ
db\sp a_u$. Stratonovich integrals have much better
transformation
properties than \Ito\ integrals, and are thus useful in a
geometric
context.

The covariance of equation (\ref{ELW})
allows one to solve the equation globally on the manifold
(although
quite sophisticated patching techniques between different
coordinate
patches are required). We have seen above that a solution to a
stochastic differential equation provides us with a Feynman-
Kac formula
for a Hamiltonian which may be deduced from the stochastic
differential
equation. In the case of the stochastic differential equation
(\ref{ELW}) the corresponding Feynman-Kac formula is

\Beq
\exp(-Hs)f(x) = \Ebold f(x_s)
\Eeq
with Hamiltonian
\Beqa
H &=& -\Half V\sp i_a V\sp j_a \partial_i \partial_j - \Half V\sp j_a
\partial_j
V\sp i_a \partial_i \End
&=& - \Half V_aV_a
\Eeqa
This Hamiltonian is globally defined, as one would expect from
the
covariance of the corresponding stochastic differential
equation.

The second component in the theory of geometric Brownian paths
on an
$m$-dimensional Riemannian manifold $M$ is the existence of a
canonical
set of vector fields $V_a,a=1, \dots ,m$ on $O(M)$, the bundle
of
orthonormal frames on $M$. In local coordinates about the
point
$(x,e_a)$ of $O(M)$,
\Beq
V_a = e\sp i_a \partial_i  -e\sp i_ae\sp j_b\Con jki
\frac{\partial}{\partial
e\sp k_b}.
\Eeq
These vector fields lead, by the process described above, to a
Feynman-Kac formula for the Hamiltonian
\Beq
H = -\Half V_a V_a
\Eeq
which is the scalar Laplacian when applied to functions on
$O(M)$ which
depend on $x$ but not
on $e_a$, that is, to functions on $M$.

By using fermionic paths (in addition to bosonic ones) on a
carefully
constructed supermanifold, it is possible to extend this
approach to
obtain a Feynman-Kac formula for the Laplace-Beltrami operator
$L =
\Lbe$ on the space of forms on a Riemannian manifold, and to
the twisted Laplace-Beltrami operator on forms which take
their values
in a vector bundle over the manifold.

Given an $m$-dimensional Riemannian manifold $M$, together
with an
$n$-dimensional Hermitian vector bundle $E$ over $M$, the
required
supermanifold $S(E)$ has dimension $(m,m+n)$ and is
constructed in the
following manner. Suppose that $\{U_{\alpha}\}$ is an open
cover of $M$
by coordinate neighbourhoods which are also trivialisation
neighbourhoods of the bundle $E$, with
\Beq
h_{\alpha\beta}:U_{\alpha} \cap U_{\beta} \to U(n)
\Eeq
the transition functions of the bundle. Then the required
supermanifold
has local coordinates $x\sp i_{(\alpha)},i=1, \dots, m ,
\theta\sp i_{(\alpha)}, i=1, \dots , m ,\eta\sp p_{(\alpha)}, p=1,
\dots, n$
with coordinate changes on overlapping neighbourhoods being as
on the
underlying manifold for the even coordinates $x\sp i$ and, for
the odd
coordinates,
\Beqa
\theta\sp i_{(\beta)} &=& \frac{\partial x\sp i_{(\beta)}}{\partial
x\sp j_{(\alpha)}} \theta\sp j_{(\alpha)} \End
\eta\sp p_{(\beta)} &=& h_{\alpha\beta}{}\sp p{}_q (x_{(\alpha)})
\eta\sp q_{(\alpha)}.
\Eeqa
(Full details of the patching construction of this
supermanifold from
its transition functions may be found in \cite{ERICE}.)

Now functions on this supermanifold of the form
\Beq
f(x,\theta,\eta) = \sum_{p=1}\sp n \sum_{\mu \in M_n} f_{\mu
p}(x)
\theta\sp {\mu} \eta\sp p
\label{FUN}\Eeq
correspond to twisted forms on $M$ with $\theta\sp {\mu}
\leftrightarrow
dx\sp {\mu}$, and the action of a $U(n)$ matrix $(A\sp p{}_q)$
represented by
$A\sp p{}_q \eta\sp q\frac{\partial}{\partial\eta\sp p}$. The Laplace-
Beltrami
operator $\Lbe$ can thus be expressed  as a differential
operator on
this space. In fact one can extend the proof of the \Wbk\
formula given
by Cycon, Froese, Kirsch and Simon  \cite{Simetal} to obtain
the twisted
\Wbk\ formula
\Beqa
\Lbe &=& - \Half (B - R\sp j_i(x)\theta\sp i\Del j - \Half \Cur kijl
(x)
\theta\sp i \theta\sp k \delta_j \delta_l \End
 &+& \Frac14 [\psi\sp i,\psi\sp j] \Cuf ijpq (x) \eta\sp p \Det q),
\Eeqa
where $B$ is the twisted Bochner Laplacian
\Beq
B = g\sp {ij} ( D_iD_j - \Con ijk D_k )
\Eeq
with
\Beq
 D_i = \partial_i + \Con ijk (x) \theta\sp j \Del k + \Cog irs
\eta\sp r \Det
s,
\Eeq
and $\Del i = \partial/\partial\theta\sp i,\Det p =
\partial/\partial\eta\sp p$ and $\psi\sp i = \theta\sp i +
g\sp {ij}(x)\Del j$.

Now let $\Som$ denote the supermanifold obtained by including
even
coordinates in $S(E)$ so that the underlying even manifold is
$O(M)$,
and consider the vector fields $W_a, a=1, \dots ,m$ on this
supermanifold where
\Beq
W_a = e\sp i_a \partial_i - e\sp j_a e\sp k_b \Con jki
\frac{\partial}{\partial
e\sp i_b} -e\sp j_a \theta\sp k \Con jki \Del i - e\sp j_a \eta\sp r \Cog
jrs
\Det s.
\Eeq
Because of the particular nature of the transition functions
of $\Som$
this defines $m$ vector fields globally on $\Som$. Now
\Beq
W_aW_a = B,
\Eeq
and thus, if we find stochastic processes
$x\sp i_s,\xi\sp i_s,e\sp i_{a,s},
\eta\sp p_s, i,a=1 , \dots , m, p=1, \dots , n$ which satisfy
\Beqa
x\sp i_s &=& x + \int_0\sp s e\sp i_{a,u} \circ db\sp u \End
e\sp i_{a,s} &=& e\sp i_a + \int_0\sp s - e\sp l_{a,u} \Con kli (x_u)
e\sp k_{b,u}
\circ db\sp b_u \End
\xi\sp i_s &=& \theta\sp i + \theta\sp a e\sp i_{a,s} + \int_0\sp s \Lba -
\xi\sp j_u \Con
jki (x_u) e\sp k_{b,u} \circ db\sp b_u \End
& &- \theta\sp a_u de\sp i_{a,u} + \Frac i4 \xi\sp j_u
R\sp i{}_{jk\ell}(x_u)
\xi\sp k_u \pi\sp {\ell}_u du \Rba \End
\eta\sp p_s &=& \eta\sp p + \int_0\sp s \Lba -e\sp j_{a,u} \eta\sp q_u \Cog
jqp (x_u)
\circ db\sp a_u \End
& & + \Frac14 \eta\sp q_u(\xi\sp i_u+i\pi\sp i_u)(\xi\sp j_u+i\pi\sp j_u)
\Cuf
ijqp(x_u) du \Rba,
\label{SDE}\Eeqa
where
\Beq
\pi\sp i_s = e\sp i_{a,s}\rho\sp a_s,
\Eeq
then one obtains the Feynman-Kac formula
\Beq
\exp-(Ls)  f(x,\theta,\eta) = \Ebold \Lsa f(x_s, \xi_s,
\eta_s) \Rsa
\label{FKSUP}\Eeq
where $L$ is the twisted Laplace-Beltrami operator $\Lbe$ and
$f$ is a
function of the form (\ref{FUN}).
\section{Applications to geometry}
As an illustration of the analytic power of this techniques,
the
Atiyah-Singer index theorem for the twisted Hirzebruch
signature theorem
will now be proved. (It was shown by Atiyah, Bott and Patodi
\cite{AtiBot} that the full theorem follows from this special
case by
K-theoretic arguments.)
The proof is a rigorous version of the supersymmetric proofs
of the
index theorem introduced by Alvarez-Gaum\'e \cite{Alvare} and
by Friedan
and Windey \cite{FriWin}. Various
other authors have used probabilistic methods  to prove the
index
theorem
\cite{Bismut,Elwort2,Leandr,Lott,Watana}, but these works do
not use the
fermionic paths described in this paper, which are a rigorous
version of
the paths used in \cite{Alvare} and \cite{FriWin}.

The proof makes use of the McKean and Singer formula for the
index $I$
of the twisted Hirzebruch signature complex \cite{McKSin}
\Beq
I = \Trace \gamma\sp 5 \exp (-Lt),
\Eeq
where $L$ is the Laplace-Beltrami operator $\Lbe$ as before,
and
$\Trace$ denotes a full trace over both operators and
matrices.
The result to be proved is that
\Beq
I = \int_M \Lsc \Indint  \Rsc ,
\Eeq
where $F$ is the curvature two-form of the connection $A$ on
the bundle
$E$, $\Omega$ is the curvature connection of the Riemannian
connection on $(M,g)$, and the square brackets indicate
projection onto
the $m$-form component. In fact a stronger, local result will
be proved,
that is, it will be shown that at each point $p$ of the
manifold $M$
\Beq
\lim_{t \to 0}  \trace \gamma\sp 5\exp(-Lt)(p,p) = \Lsc \Indint
\Rsc_p
\label{AST}\Eeq
where $\trace$ denotes a matrix trace.

To evaluate the left hand side of (\ref{AST}), the Feynman-Kac
formula
(\ref{FKSUP}) must be used. Addiditionally, several steps in
the
analysis of Grassmann variables are required, and Du Hamel's
formula
must be
used to extract information about the kernel of $\exp(-Lt)$
from
information about the action of the operator on functions. The
most important difference
between the approach used in heuristic physicists'
calculations of path
integrals in curved space and the approach used here is that
stochastic
differential equations are used, which means that different
paths are
used but the measure on path space is unchanged, whereas in
the
heuristic approach it is the measure which is changed, which
is a much
harder task to handle analytically.

Beginning with the anticommuting variables, the first step is
to observe
that the action of $\gamma\sp 5$ on twisted forms is
\Beq
\gamma\sp 5 \Sumsum f_{\mu p}(x)\theta\sp {\mu}\eta\sp p = \Sumsum \int
d\sp m\rho
\frac{1}{\sqrt{\det g\sp {ij}}} \exp i\rho\sp i\theta\sp j g_{ij}(x)
f_{\mu p}(x)\rho\sp {\mu}\eta\sp p.
\label{GAM}\Eeq
The operator $\gamma\sp 5$ is an involution, which resembles the
Hodge star
operator.

As we are working at one particular point $p\in M$ we will use
normal
coordinates $x\sp i$ about this point, so that $p$ has
coordintaes 0, while
\Beqa
g\sp {ij}(x) &=& \delta\sp {ij} - \Frac13 x\sp k x\sp {\ell}
R_k{}\sp i{}{\ell}{}\sp j(0)
+ \Hot \End
\det(g_{ij}(x)) &=& 1 + \Hot\End
\Con ijk &=& \Frac13 x\sp {\ell}(\Cur {\ell}jik(0) + \Cur
{\ell}ijk(0))
+ \Hot\End
\Cog irs &=& -\Half x\sp j \Cuf ijrs(0)+ \Hot
\Eeqa
It is shown by Cycon, Froese, Kirsch and Simon \cite{Simetal}
that for
the purposes of calculating $\lim_{t\to 0} \trace \gamma\sp 5
\exp(-
Lt)(p,p)$ one may replace the manifold
$M$ with $\Real\sp m$ and the bundle $E$ over $M$ with the
trivial bundle
$\Real\sp m \times \Comp\sp n$ provided that the metric and
connection are
chosen to agree with those on the manifold on a neighbourhood
of $p$ (indentified with a neighbourhood of $0$ in $\Real\sp m$
by the coordinate functions) and
to be the Euclidean metric and the zero connection outside
some compact
region of $\Real\sp m$. Thus from now on these replacements will
be made.

Using the expression (\ref{GAM}) for the action of $\gamma\sp 5$,
together with the expression (\ref{TRACE}) for the trace of an
operator
in terms of a Berezin integral, one finds that
\Beq
\trace \gamma\sp 5 \exp (-Lt)(0,0)
= \int d\sp m\rho d\sp m\theta d\sp n\eta \exp (-Lt) (0,0,\rho,-
\theta,\eta,-
\eta) \exp
i \rho.\theta\,.
\Eeq
Now the Feynman-Kac formula (\ref{FKSUP}) tells us about the
action of
$\exp(-Lt)$ on functions. Using Du Hamel's formula
\cite{Getzle} one may
extract information about the kernel of this operator. Suppose
that
$H_0$ is the operator $-\Half\partial_i\partial_i$ on $\Cinf
C(\Real\sp m)$. Then Du Hamel's formula states that
\Beqa
\lefteqn{\exp(-Lt)(x,x',\theta,\theta',\eta,\eta') - \exp(-
H_0t)
(x,x',\theta,\theta',\eta,\eta')=}\End
& & \int_0\sp t \exp(-L(t-s))(L-H_0)\Lsa \exp-
H_0s(x,x',\theta,\theta',\eta,\eta')\Rsa,
\Eeqa
where the operators $L$ and $H_0$ act with respect to the
variable
$x,\theta$ and $\eta$.
Taking the supertrace of both sides (that is, operating with
$\gamma\sp 5$
and taking the trace) one finds that
\Beqa
& &\trace \gamma\sp 5 \exp(-Lt)(0,0) = \int d\sp m\theta d\sp m\rho
d\sp n\eta
\int_0\sp t \Lsb
\exp(-L(t-s))(L - H_0) \End
& & \times\Lbb \exp(-H_0s(x,0,\rho,\theta,\eta,\pi))\Rbb
|_{x=0,\pi = - \eta}\exp-\theta.\rho \Rsb,
\Eeqa
(using the fact that $\trace \gamma\sp 5 \exp(-Lt)=0$).

Now
\Beqa
& &\exp(-H_0t)(x,0,\theta,\theta',\eta,\eta')
= \int d\sp m\rho d\sp n\kappa \End
& &\Factor t\exp(\frac{x\sp 2}{2t}) \exp-i\rho.(\theta-
\theta') \exp-i\kappa(\eta-\eta')
\Eeqa
so that (using the Feynman-Kac formula (\ref{FKSUP}), and
carrying out
some integration)
\Beqa
& &\trace \gamma\sp 5 \exp(-Lt) (0,0) = \Ebold \Lsc\int d\sp m
\theta d\sp n\eta
d\sp n\kappa \End
& &\int_0\sp t ds \Factor s F_s(x_{t-s},\xi_{t-s},\theta,\eta_{t-
s},\kappa)
\exp-\kappa.\eta\Rsc,
\Eeqa
where
\Beq
F_s(x,\theta,\rho,\eta,\kappa) = (L - H_0) (\exp \frac{-
x\sp 2}{2s}\exp-
i\rho.\theta \exp-i\kappa.\eta).
\Eeq
Now this can be estimated as $t$ tends to $0$ in the following
way.
First let $\One x\sp i_s,\One{\xi}\sp i_s,\One{\eta}\sp p_s$
satisfy the stochastic differential equations
\Beqa
\One x\sp i_s &=& b\sp i_s \End
\One{\xi}\sp i_s &=& \theta\sp i + \theta\sp a_s\delta_a\sp i + \int_0\sp s
(\Frac13
\One{\xi}\sp j_u\One x\sp {\ell}_u(\Cur {\ell}kji + \Cur
{\ell}jki)db\sp k_u \End
& & + \Frac13 \One{\xi}\sp j_uR\sp i_j du - \Frac i4
\One{\xi}\sp j_u\One{\xi}\sp k_u
\rho\sp {\ell}_u R_{jk}{}\sp i{}_{\ell} du) \End
\One{\eta}\sp p_s &=& \eta\sp p + \int_0\sp s \Frac14 \One{\eta}\sp q_u
(\theta\sp i_u-i\pi\sp i_u)(\theta\sp j_u-i\pi\sp j_u)\Cuf ijqp du.
\label{SDESIM}\Eeqa
(Here $\Cur ijkl = \Cur ijkl(0)$, and indices are raised and
lowered by
$g\sp {ij}(0) = \delta\sp {ij}$.)
Now solutions to this set of stochastic differential equations
are close
to
those of (\ref{SDE}). In fact $x_s - \One{x}_s \sim
\sqrt{s\sp 3}, \xi_s-
\One{\xi}_s \sim s, \eta_s -\One{\eta}_s \sim s$ and
$e\sp i_{a,s} -
\delta\sp i_a \sim s$ \cite{SCSTWO}. Thus
\Beq
\lim_{t \to 0} \trace\gamma\sp 5\Lbb\exp(-Lt)(0,0) - \exp(-
\One Lt)(0,0)\Rbb
= 0
\Eeq
where $\One L$ is the Hamiltonian corresponding to the
simplified
stochastic differential equations (\ref{SDESIM}).

This simpler Hamiltonian $\One L$  can be handled by flat
space path
integral
methods. Use of these, together with further use of Du Hamel's
formula,
shows that
\Beqa
& &\lim_{t \to 0} \trace \gamma\sp 5 \exp (-\One L t)(0,0)\End
&=&\lim_{t \to 0}\int d\sp m\rho d\sp m\theta d\sp n\eta \exp(-\One L
t)(0,0,\rho,-
\theta,\eta,-\eta) \exp i \rho.\theta
\End
&=&\lim_{t \to 0}\int d\sp m\rho d\sp m\theta d\sp n\eta \exp(-\Two L
t)(0,0,\rho,-
\theta,\eta,-\eta) \exp i \rho.\theta
\Eeqa
where
\Beq
\Two L = H_{x} + H_{\theta} + H_{\eta}
\Eeq
with
\Beqa
H_x &=&-\Lbb\Half \partial_i\partial_i +\Half \Xhat{\ell}
\Phitt j\Phitt i
\Cur ji{\ell}k \partial_k\End
& &\qquad+{\textstyle{1\over8}}\Xhat k\Xhat{\ell}{\phi\sp i\phi\sp j
\phi\sp {n'}\phi\sp {m'}\over (2\pi t)\sp 2}
R_{n'ikp} \Cur {m'}j{\ell}p\Rbb,\End
H_{\theta} &=& -{1\over4}R_{ijk\ell}{\phi\sp i\phi\sp j\over2\pi t}
\psi\sp k \psi\sp {\ell}\quad{\rm and} \End
H_{\eta}&=&-\Phitt i\Phitt j \Cuf ijpq \Etah
p\Det q,
\Eeqa
with $\phi = \sqrt t \theta$. (This rescaling allows one to
pick out
those terms which survive in the limit as $t$ tends to zero.
Note that
$d\theta = \sqrt t d\phi$.)
Now $\exp-H_x t(0,0)$ can be evaluated using the result given
by
Simon \cite{Simon} for $\Real\sp 2$ that, if
\Beq
H= -\Half\partial_i\partial_i +
{iB\over{\textstyle2}}(x\sp 1\partial_2-
x\sp 2\partial_1) + {\textstyle{1\over8}}B\sp 2((x\sp 1)\sp 2+(x\sp 2)\sp
2),
\Eeq
then
\Beq
\exp-Ht(0,0) = {B\over 4\pi\sinh(\Half Bt)}.
\Eeq
Thus, if $\Curv kl = \Half\phi\sp i\phi\sp j \Cur ijkl$ is regarded
as an
$m\times m$ matrix, skew-diagonalised as
\Beqa
\Lba\Omega_k{}\sp l\Rba = \left(\matrix
{0 &\Omega_1 &\ldots&0 &0\cr
 -\Omega_1 &0&\ldots&0 &0\cr
\vdots&\vdots&\ddots&\vdots&\vdots\cr
0&0&\ldots&0&\Omega_{\Half m}\cr
0&0&\ldots&-\Omega_{\Half m}&0\cr}\right),
\Eeqa
then
\Beq
\exp-H_x t(0,0)
= \prod_{k=1}\sp {m/2} {i\Omega_k\over2\pi t}\,
{1\over\sinh({i\Omega_k\over2\pi})}.
\Eeq
Also, using fermion paths \cite{GBM} or direct calculation,
\Beqa
& &\exp-tH_{\theta}(\theta,\theta') \End
& &= \int d\sp m\rho \Lbb\exp -
i\rho(\theta-\theta')\End
& & \qquad\prod_{k=1}\sp {m/2} \Lba \cosh {i\Omega_k\over2\pi}
+ (\theta\sp {2k-1}
+i\rho\sp {2k-1})(\theta\sp {2k}+i\rho\sp {2k}) \sinh
{i\Omega_k\over2\pi}\Rba\Rbb.
\Eeqa
Thus
\Beqa
& &\trace \gamma\sp 5 \exp-H t(0,0)\End
&=&\int d\sp m\phi \prod_{k=1}\sp {m/2} {i\Omega_k\over2\pi}\,
{1\over\sinh({i\Omega_k\over2\pi})} \cosh {i\Omega_k\over4\pi}
{\rm tr} (\exp -\phi\sp i\phi\sp j {F_{ij}\over2\pi}).
\Eeqa
Hence
\Beq
\trace \gamma\sp 5 \exp -Lt(p,p)
=\Biggl[{\rm tr}\,
\exp\Lbc {-F\over2\pi}\Rbc \det \Lbc {i\Omega/2\pi\over\tanh
i\Omega/2\pi}\Rbc\sp {\Half}\Biggr]\Biggr|_p
\Eeq
as required.

\end{document}